\documentclass[12pt]{article}
\usepackage{latexsym}
\usepackage{epsfig}

\usepackage{graphicx}
\usepackage{epstopdf}

\usepackage{epsfig,amssymb,amsmath,amscd}

\newcommand{\bq}{\begin{equation}}
\newcommand{\eq}{\end{equation}}
\newcommand{\bqa}{\begin{eqnarray}}
\newcommand{\eqa}{\end{eqnarray}}
\newcommand{\ben}{\begin{enumerate}}
\newcommand{\een}{\end{enumerate}}
\newcommand{\bc}{\begin{center}}
\newcommand{\ec}{\end{center}}
\newcommand{\bqb}{\begin{eqnarray*}}
\newcommand{\eqb}{\end{eqnarray*}}

\def\pr#1#2#3{Phys. Rev. ${\bf{#1}}$, #2 (#3)}

\def\pl#1#2#3{Phys. Lett. ${\bf{#1}}$, #2 (#3)}

\def\np#1#2#3{Nucl. Phys. ${\bf{#1}}$, #2 (#3)}

\def\epj#1#2#3{Eur. Phys. J. ${\bf{#1}}$, #2 (#3)}

\def\jmp#1#2#3{J. Mod. Phys. ${\bf{#1}}$, #2 (#3)}

\begin{document}
\pagenumbering{arabic}
\thispagestyle{empty}
\def\thefootnote{\fnsymbol{footnote}}
\setcounter{footnote}{1}

\begin{flushright}
August 3, 2017\\
 \end{flushright}

\begin{center}
{\Large {\bf Overview of the CSM concept and its tests}}.\\
 \vspace{1cm}
{\large F.M. Renard}\\
\vspace{0.2cm}
Laboratoire Univers et Particules de Montpellier,
UMR 5299\\
Universit\'{e} Montpellier II, Place Eug\`{e}ne Bataillon CC072\\
 F-34095 Montpellier Cedex 5, France.\\
\end{center}

\vspace*{1.cm}
\begin{center}
{\bf Abstract}
\end{center}

We recall the motivations for the Compositeness Standard Model(CSM) concept, 
its precise description, the procedures for its applications and the 
particular constraints that it requires. We present its most spectacular 
predictions for typical processes observable at present and future colliders. 
To the previous results we add the treatment of the inclusive 
processes $e^+e^- \to H,Z,W~or~t+ anything$.

\vspace{0.5cm}
PACS numbers:  12.15.-y, 12.60.-i, 14.80.-j;   Composite models\\

\def\thefootnote{\arabic{footnote}}
\setcounter{footnote}{0}
\clearpage

\section{INTRODUCTION}

In spite of its success the SM is not totally satisfactory. 
Various BSM possibilities have been proposed
including SM extensions, supersymmetry, additional
strong sectors,...\\
In most of these proposals the number of free parameters, 
of basic states and interactions increases in some cases enormously. 
No experimental indication for some direction of search has been
found up to now.  Maybe the corresponding new physics scale lies beyond the
experimental reach. In any case introducing a large number of new states and 
interactions would generate an increasing number of questions about their origin.\\
\underline{Why compositeness?}\\
We would prefer that new physics develops towards simplicity. Traditionally
compositeness may be such a way with constituents named preons, subquarks,...,
see for example \cite{comp}.\\
There is however no special experimental indication for this possibility, 
maybe again due to a very high substructure scale?\\
Although no dynamical model with computational possibilities
is available yet we want to believe that compositeness is still conceivable.\\
The heavy top quark has been a new motivation for full or partial top compositeness
\cite{partialcomp} and for the addition of Higgs compositeness \cite{Hcomp2,
Hcomp3,Hcomp4}.\\
Other states could still be elementary or
partially composite with a very small mixing effect.\\
\underline{Why CSM?}\\
In spite of its lack, SM has important and efficient structures
in the gauge and Higgs sectors that one may want to maintain
(structures of the gauge and Higgs couplings, Goldstone 
equivalence with longitudinal W,Z components).
The Composite Standard Model (CSM) concept consists in assuming that the 
compositeness picture preserves these SM structures.
We have yet no precise model to propose from which one may compute
the observables and check these properties. We assume that these SM structures
are reproduced in an effective way at low energies and that the first
consequence of compositeness (spatial extension) will be the occurence of 
form factors affecting the concerned basic couplings.\\
In this paper we review the results of preliminary works concerning CSM and 
we add some new studies
(in particular for inclusive processes)
which compare CSM conserving and CSM violating predictions
for various processes in $e^+e^-$, $\gamma\gamma$ and hadronic 
collisions.\\

Contents: In Sect.2 we recall the basis of a CSM description.
In Sect.3 we develop the 3-step strategy
with (1) the detection of the presence of form factors, (2) the check that they satisfy
the CSM constraints, (3) its confirmation with more involved processes. The summary and the possibility of future developments are presented in Sect.4.\\

\section{CSM description}

We have established an effective description of substructure effects
with what we call the CSM concept. It consists in assuming that
the pure SM is preserved at low energy with its usual set of basic
couplings. We have no precise model allowing a direct computation of the CSM observable
effects. But with this concept
no anomalous coupling creating immediate deviation from SM should appear.
The spatial extension due to compositeness
would only generate an energy dependence of the point-like couplings
which means a form factor affecting them, but being
close to 1 at low energy, and controlled at high energy by a
new physics scale related to the binding of the constituents.\\
An example of such form factor that we will use 
in our illustrations is:
\bq
F(s)={s_0+M^2\over s+M^2}~~\label{FF}
\eq
\noindent
with the new physics scale $M$ taken for example in the few TeV range.\\
It will be applied 
to the top quark ($t_L$ and/or $t_R$), to the Higgs boson
and to the complete Higgs doublet with the Goldstone equivalence with $W_L,Z_L$.\\
The preservation of the SM structure  will require some relation, that we
call a CSM constraint, between
the form factors affecting the different basic couplings. 
A typical case is given by the famous 
cancellations ensuring a good high energy behaviour of the amplitudes
involving longitudinal gauge bosons which is preserved when the involved
form factors satisfy the CSM constraint.\\

The strategy for establishing a CSM analysis should then proceed as follows:\\
1) detect the presence of compositeness form factors,\\
2) check if they satisfy CSM constraints in basic processes,\\
3) confirm the expected consequences for more involved processes.\\

\section{Three step strategy}

\subsection{1st step: detect the presence of form factors}

In this first step we are looking for the presence of form factors
in Higgs boson couplings and in top quark couplings.\\

a)  Higgs form factor\\

There are no basic $\gamma HH$ nor $ZHH$ coupling allowing to define a
pure Higgs boson form factor.
The simplest place concerning a pure $H$ form factor would be the $HHH$ coupling,
not for looking for anomalous components, but for an s-dependence when
one $H$ line is off-shell; this is difficult to measure, see \cite{mumuH}.\\

One can then look for the existence of a $ZZH$ form factor, see \cite{WLZL}.
However if compositeness preserves the whole SM Higgs doublet structure then
form factors may affect the 
$ZG^0H$, $W^{\pm}G^{\pm}G^0$, $W^{\pm}G^{\pm}H$, $HG^{\pm,0}G^{\pm,0}$, 
$ZG^{\pm}G^{\pm}$, $\gamma G^{\pm}G^{\pm}$ 
couplings and, if the equivalence is preserved as assumed by CSM, the corresponding form factors
when  $G^{\pm,0}$ are replaced by $W^{\pm}_L$, $Z_L$.
So one can check the SU(2)*U(1) structure of these form factors, which means to check
if the presence of a Higgs form factor can be generalized to a set of $G^{\pm,0}$ form factors 
transmitted to $W^{\pm}_L$, $Z_L$.\\ 

The first study could be the observation of a $ZZ_LH$ form factor
equivalent to a $ZG^0H$ one in the $e^+e^-\to ZH$ process, see \cite{WLZL}.\\
In Fig.1 we have illustrated the consequences for the $e^+e^-\to ZH$ cross section
of the presence of such form factor in the case of pure $Z_L$ and of unpolarized
$Z$ production. A direct measurement of this form factor can indeed easily 
be done in this channel.\\

In principle a second study could consist in checking the presence 
of $\gamma WW$, $\gamma\gamma WW$ form
factors in the $\gamma\gamma \to WW$ process which is 
simpler than $e^+e^- \to WW$ that we will consider later on below because here no special $W_L$ cancellation occurs.\\

In Fig.2 one sees that in order to get a clear signal from
the measurement of the $\gamma\gamma \to WW$ cross section (dominated by $W^+_TW^-_T$
production if no form factor affects the transverse $W_T$ states) one should
detect and restrict the analysis to the pure $W^+_LW^-_L$ final state.\\

b)  Top quark form factor\\

The simplest process for looking for the presence of left and right top quark form factors 
is $e^+e^-\to t\bar t$.
One can obviously detect the presence of 
$\gamma t_{L,R}t_{L,R}$ and $Zt_{L,R}t_{L,R}$ form factors with the
options of both  $t_{L,R}$ or of only $t_{R}$ compositeness.\\
The modification of the energy dependence of the cross section would directly
measure the size of the corresponding form factors.\\ 
See \cite{trcomp,tLR} for illustrations iin particular in the case of pure $t_{R}$ compositeness.
It is also shown how
the processes $gg \to t\bar t$ and $\gamma\gamma \to t\bar t$ should confirm
these informations.\\

\subsection{2nd step: CSM constraints}

We first consider the case of pure Higgs compositeness and then 
the case where both the Higgs boson and the top quark are composite.\\

\underline{CSM constraint in the pure gauge-Higgs sector and Goldstone equivalence}\\ 

One wants to check if the form factors of the Higgs sector satisfy the CSM properties related to
the gauge structure and the Goldstone equivalence.\\

The process $e^+e^-\to W^+W^-$ provides a basic case for testing these
properties due to the presence of important cancellations among
different parts of its amplitudes.

At Born level it is described with 2 types of diagrams, neutrino exchange 
with $We\nu$ coupling and
$\gamma,Z$ exchange with $\gamma,Z-WW$ coupling. The well-known SM gauge feature is
the cancellation of these 2 types of contributions for the $e^+_Re^-_L\to W^+_LW^-_L$
amplitude which would otherwise increase with energy and violate unitarity. 
Another SM aspect is the equivalence with $e^+e^-\to G^+G^-$ whose energy dependence is automatically well behaved.\\
We have checked that the introduction of an (even minor) form factor in the 
$\gamma,Z-W^+_LW^-_L$ coupling immediately destroys this cancellation and leads
to an inacceptable increase of the $e^+_Re^-_L\to W^+_LW^-_L$ amplitude (see  ref.\cite{WLZL}).
If the equivalence with $e^+_Re^-_L\to G^+_LG^-_L$ is maintained, as assumed by
CSM, the introduction of the same form factor for the $\gamma,Z-G^+G^-$ coupling
leads to a well-behaved acceptable amplitude (see \cite{WLZL}).\\
On another hand there is no such problem for the $e^+_Le^-_R\to W^+_LW^-_L$ amplitude
which receives no contribution from neutrino exchange and is well-behaved as long
as the $e^+_Le^-_R\to \gamma \to W^+_LW^-_L$ and $e^+_Le^-_R\to Z \to W^+_LW^-_L$
contributions combine properly as in the SM case, which means the same form
factor effects in the $\gamma-W^+_LW^-_L$ and $Z-W^+_LW^-_L$ couplings. Also one
can check that the equivalence with the $e^+_Le^-_R\to \gamma,Z \to G^+G^-$
amplitude is quickly satisfied at high energy when the same form factor is 
applied (see \cite{WLZL}).\\

In a first study one may assume that Goldstone equivalence is preserved in some
effective manner by CSM. We will call this possibility as CSMG.\\
So we will compare two different situations,
one without the equivalence requirement (CSMFF) with arbitrary form factors
affecting longitudinal gauge bosons
and one assuming that the Goldstone equivalence is preserved by the CSM picture
(CSMGFF) in some effective manner  and that similar arbitrary form factors
affect the corresponding Goldstone bosons couplings.\\
This ensures a good 
high energy behaviour (even with new effects) such that the presence 
of form factors produces immediately a decreasing effect.

This assumption is applied to  $e^+e^-\to W^+W^-$ in Fig.3.
The upper panel concerns the unpolarized $e^+e^-$ case
with either a crude $W_L$ form factor (LFF) which violate the CSM constraint
and generates a strong effect because of the violation of
the usual SM cancellation,
or with the assumption of CSM Goldstone equivalence (CSMGFF)
which keeps the cancellation and leads to a normal decrease.
In the lower panel, one makes the comparison with the cases of $e^-_R$
polarization and again either a crude $W_L$ form factor (LFF) or with
the CSM Goldstone equivalence(CSMGFF), which are now both acceptable
because no (preserved or violated) cancellation is present.\\

One can pursue this type of study with other processes involving Higgs bosons and/or
longitudinal gauge bosons and/or top quarks.\\
A first set corresponds to the famous WW, WZ, ZZ scattering occuring through
gauge and Higgs bosons exchanges. Introducing form factors for $H$, $W_L$ and $Z_L$
couplings, the cancellation would also require that they have a common $F(s)$ shape.
This is the crudest situation, but assuming that CSM will provide
in some effective way the equivalence such that one can replace $W^{\pm}_L$, $Z_L$ by
$G^{\pm,0}$ and then apply the form factors (CSMG picture) will give immediately
a correct behaviour.\\

\underline{CSM constraint for both Higgs and top compositeness}\\

We can apply this discussion to any $W^-_L$ production process,
for example obviously to $e^+e^-\to WWH, ZZH$, but also to other processes
involving the top quark.
In any $W^-$ production process, because of the presence of different
types of diagrams,
the complicated procedure of cancellation of the  strongly
increasing and unitarity violating $W^-_L$ component 
plays an important role in the resulting effects of the  
form factors.
For example the fact that the top quark may be composite and
have a form factor whereas the bottom quark remains elementary
without form factor creates an importent lack of cancellation.
The sizes of the effects are then specific to each process
because on the one hand form factors lead to a decrease with the energy
whereas on the other hand the lack of cancellation leads to an increase.
So we will again compare for each process the two different situations,
CSM without the equivalence requirement and arbitrary form factors
affecting longitudinal gauge bosons
and CSMG with the Goldstone equivalence and similar arbitrary form factors
affecting the corresponding Goldstone bosons couplings.\\
For $W^-_L$ production processes involving also the top quark we 
will show the differences between CSMtLR and CSMGtLR in the case of both
$t_{L,R}$ compositeness and those 
between CSMtR and CSMGtR assumptions in the case of pure
$t_{R}$ compositeness.\\

On another hand one may want to check if possible top quark form factors 
are in some way related to the ones of the Higgs sector; this could
be natural if the top quark and the Higgs boson have the same subconstituents. 
In fact we have seen that such relations may be imposed by the energy behaviour 
of the $ZH$  one loop production amplitudes
in $gg$ and $\gamma\gamma$ collisions. 
In \cite{CSMgg,CSMgamgam} we have remarked that the $gg\to ZH$ and $\gamma\gamma\to ZH$
processes are particularly sensitive to the presence of form factors because they could 
destroy a peculiar SM cancellation between diagrams involving Higgs boson and top quark
couplings, top loops and $G^0$ exchange in the s-channel.\\
But we have also shown that this cancellation can be preserved provided 
a special relation between form factors is satisfied. We considered this
relation as a specific CSM property.
Introducing five arbitrary effective form factors chosen as
$F_{G^0Z_LH}(s)=F_{ZZ_LH}(s)$, $F_{Htt}(s)$, $F_{Gtt}(s)$, 
$F_{tR}(s)$, $F_{tL}(s)$ this preservation occurs provided the following 
CSM constraint is satisfied:
\bq
F_{G^0Z_LH}(s)F_{Gtt}(s)(g^Z_{tR}-g^Z_{tL})=
F_{Htt}(s)(g^Z_{tR}F_{tR}(s)-g^Z_{tL}F_{tL}(s))~~\label{CSMconsZH}
\eq
The same constraint can be inferred by looking directly at the high energy behaviour 
of the $t\bar t\to Z_LH$ amplitudes.\\

The  above procedure can be generalized to $t\bar t$ production amplitudes in ZZ 
and WW collisions \cite{ZZWWtt}, especially with longitudinal Z and W which could be composite
and nevertheless preserve the Goldstone equivalence.\\ 

First the $Z_LZ_L\to t\bar t$
process gives the constraint
\bq
-{1\over2}g_{HZZ}g_{Htt}F_{HZZ}(s)F_{Htt}(s)=
m_{t}((g^Z_{tR}F_{tR}(s)-g^Z_{tL}F_{tL}(s))^2~~\label{CSMconsFFZ}
\eq
\noindent
and the $W_LW_L\to t\bar t$ processes  requires
\bq
F_{HWW}(s)F_{Htt}(s)=
F^2_{tL}(s)=F_{VWW}(s)F_{tL}(s)~~\label{CSMconsFFW}
\eq
In this second case in order to recover the SM structure one needs
to require that the $\gamma tt$ and $Ztt$ form factors are similar as well
as the $F_{\gamma WW}(s)$ and $F_{ZWW}(s)$ form factors. Finally the two
above constraints require that all the involved form factors have a common 
$F(s)$ shape.\\

\underline{An effective top mass?}\\

Instead of requiring this crude property of unique form factor there is
another way of preserving CSM.
As mentioned in \cite{trcomp}, in some processes where, after the cancellation
of the increasing contributions, the resulting SM amplitudes 
appear to be proportional to the top mass $m_t$, the alternative possibility
consists in introducing a (decreasing) effective mass $m_t(s)$, 
depending on the compositeness scale, which would finally ensure a good high energy behaviour.\\
This may be also considered as consistent with the CSM concept as it preserves the
SM structure.\\
We will illustrate the effects of this choice as compared to the other CSM and CSMv cases
in the following studies and in the next Section.\\

\underline{Application to $gg,\gamma\gamma \to ZH$}\\

Details about the behaviour of the various helicity amplitudes
can be found in \cite{CSMgg,CSMgamgam} for both proceeses and the 4 choices of form factor types,
those which violate CSM, with only Higgs form factor (CSMvH), with
both Higgs and top form factors (CSMvt), those which preserve
the above CSM constraint with both $t_L$ and $t_R$ form factors
(CSMtLR) or with only a $t_R$ form factor (CSMtR).\\
Explicitly, in the illustrations that we will present we will
use the $F(s)$ form factor of eq.(\ref{FF})
and we will make comparisons between the 4 or 6 following choices:\\

CSMtLR and CSMGtLR: $F_{t_R}(s)=F_{t_L}(s)=F(s)$
 and $F_G(s)=F_H(s)=F(s)$
 keeping the top mass at its bare value,\\
 
CSMtR and CSMGtR: $F_{t_L}(s)=1$ $F_{t_R}(s)=F(s)$
 and $F_G(s)=F_H(s)=F(s)$, with the effective top mass
 $m_t(s)=m_tF(s)$,\\
 
CSMvt: different form factors for $t_L$ (ex; M= 10 TeV)
 and for $t_R$ (ex; M= 15 TeV), and $F_G(s)=F_H(s)=F(s)$,
 with a bare top mass,\\
 
CSMvH: no top form factor but $F_G(s)=F_H(s)=F(s)$
 and the bare top mass.\\

In Fig.4 we illustrate the differences between 4 choices for the
ratios of new cross sections over the SM one in the case of the $gg\to ZH$
process. The $\gamma\gamma \to ZH$ process would give complementary
results in particular when polarized photon beams can be used, see \cite{CSMgamgam}.\\

\subsection{3rd step: Confirmation of the CSM constraints}

Careful analyses of each of the above processes should determine the detailed properties
of the $H$ and top quark compositeness and in particular to see if the
modifications of the SM predictions correspond to
CSM conservation or to CSM violation in each of these sectors.\\

The next step may consist in studying other processes in order to
check these properties.
More involved processes like
$t\bar t H$, $t\bar t Z$ and $t\bar b W$ production 
in $e^+e^-$ or in hadronic collisions and 
inclusive distributions like $e^+e^- \to H,Z,W,t+....$ should be very productive
because they directly involve
both the Higgs boson and the top quark sectors.\\

In ref.\cite{CSMee,CSMggttH} we have studied the processes 
$e^+e^-, gg, \gamma\gamma \to t\bar t H, t\bar t Z, t\bar b W^-$.
Large and specific  effects of Higgs and top quark form factors have
been found. The most spectacular ones appear (as expected from the
special longitudinal cancellations) in the $t\bar b W^-$ case.
We show them for the 3 processes $e^+e^-, gg, \gamma\gamma \to t\bar b W^-$
in Fig.5,6,7 with the 6 options of form factors.
Strong effects appear from CSM violating cases. When the Goldstone equivalence
is imposed (in $CSMG...$ cases) a reasonable form factor effect is recovered.\\

We now look at the inclusive distributions $e^+e^- \to H,Z,W,t+....$ which
should cumulate the various effects.
The illustrations are shown for the reduced momentum $x={2p\over\sqrt{s}}$ at
a center of mass energy of 4 TeV and an angle of ${\pi\over3}$.\\

\underline{Inclusive distribution $e^+e^- \to H+....$}\\

In SM the main leading channels contributing
are $Hff$(essentially $Htt$), $HHZ$, $HZZ$, $HZ\gamma$, $H\gamma\gamma$, $HWW$.
The resulting $x$ distribution  
is shown in Fig.8, showing the sensitivity to $Htt$, $Vtt$, and bosonic $H$ couplings.
With the usual 4 choices of form factors, only $t_R$ compositeness leads to small effects,
whereas larger ones appear from $H$ compositeness.\\

\underline{Inclusive distribution $e^+e^- \to Z+....$}\\

The SM channels are $Zff$ (all quarks and leptons), $ZHH$, $ZHZ$, $ZZZ$, $Z\gamma\gamma$,
 $ZZ\gamma$, $ZWW$
leading to the distributions shown in Fig.9a and Fig.9b for longitudinnally
polarized $Z_L$ and unpolarized $Z$.
The $ZWW$ contribution is particularly important due to its special SM cancellations
which can be perturbed by arbitrary form factors, especially in the $Z_L$ case.\\
Large effect are found when these cancellations are violated (CSMtLR, CSMvH,CSMvt);
they are smaller for CSMGtLR, CSMtR, CSLGtR where no violation occurs.
The effects are slightly less observable in the unpolarized $Z$ case.\\

\underline{Inclusive distribution $e^+e^- \to W+....$}\\

The SM channels are now $Wff'$ (quarks and leptons), $WWH$, $WWZ$, $WW\gamma$
and shown in Fig.10a and Fig.10b for $W_L$ and unpolarized $W$.\\
The analysis and the effects are similar to the ones in the 
inclusive $Z$ case and, with the usual 6 choices, one finds 
even more pronounced effects.\\

\underline{Inclusive distribution $e^+e^- \to t+....$}\\

The SM channels are $ttH$, $ttZ$, $ttg$, $tbW$. We consider the 3 cases with
$t_L$, $t_R$ or unpolarized $t$  for the 6 choices of form factors;
the results are shown in
Fig.11a, Fig.11b and Fig.11c respectively.\\

In the $t_L$ and unpolarized $t$ cases one can see 
large effects for CSMtLR, CSMvH, CSMvt; medium ones for CSMGtLR;
and small ones for CSMtR, CSMGtR.\\

The $t_R$ distribution can reveal different specific effects 
of  $t_R$ compositeness although the size of the effects may be weaker.\\

\subsection{The steps after....}

After having looked at all the above proceeses, if some form factor
effect is revealed, the point will be to see all of its caracteristics,
in particular if it seems to correspond to some CSM, CSMG or CSMv type.
The complete set of couplings should be tested. This will require an
amplitude analysis of all concerned processes with measurements of
angular distributions, polarization distributions, subenergy
dependences of the various cross sections.\\
The possibilities can be estimated by looking for example
for $e^+e^-$ collisions at \cite{Moortgat, Denterria, Craig, Englert}, 
for hadronic collisions at \cite{Contino,Richard} and for 
photon-photon collisions at \cite{gammagamma}.\\
Other processes, for example those involving the bottom quark,
should be studied in order to check if they are more or less
affected by compositeness.\\
All these results may then suggest some theoretical modelization of 
the underlying dynamics.\\

\section{Summary}

In this paper we have reviewed what we call the CSM concept,
which assumes that Higgs boson and top quark compositeness may
preserve the basic SM structure, including the Goldstone equivalence.\\
Describing such compositeness effects by form factors affected
to each Higgs boson and top quark coupling we have compared the
observable consequences of CSM conserving and of CSM violating
choices.\\
We have proposed a 3-step strategy for analyzing possible departures
from SM predictions in the concerned processes.\\

Step 1: detect form factors in $e^+e^-\to ZH$, in $\gamma\gamma \to WW$
and in $e^+e^-\to t\bar t$.\\

Step 2: check if special CSM constraints in $e^+e^-\to W^+W^-$ and in 
$gg,\gamma\gamma \to ZH$ are preserved.\\

Step 3: Confirm the validity of the CSM constraints in 
$t\bar t H$, $t\bar t Z$ and $t\bar b W$ production in $e^+e^-$, $\gamma\gamma$
and hadronic collisions and in inclusive processes like 
$e^+e^- \to H,Z,W,t+anything$.\\
 
Depending on the nature of the results further phenomenological and theoretical
developments may be required on the one hand for analyzing complete sets of
observables and on the other hand for establishing a dynamical description
of the effective form factors. We would like to particularly mention
the occurence of an effective top mass. This compositeness property may
suggest new ways for discussing the whole fermion spectrum.\\

\newpage

\begin{figure}[p]
\[
\epsfig{file=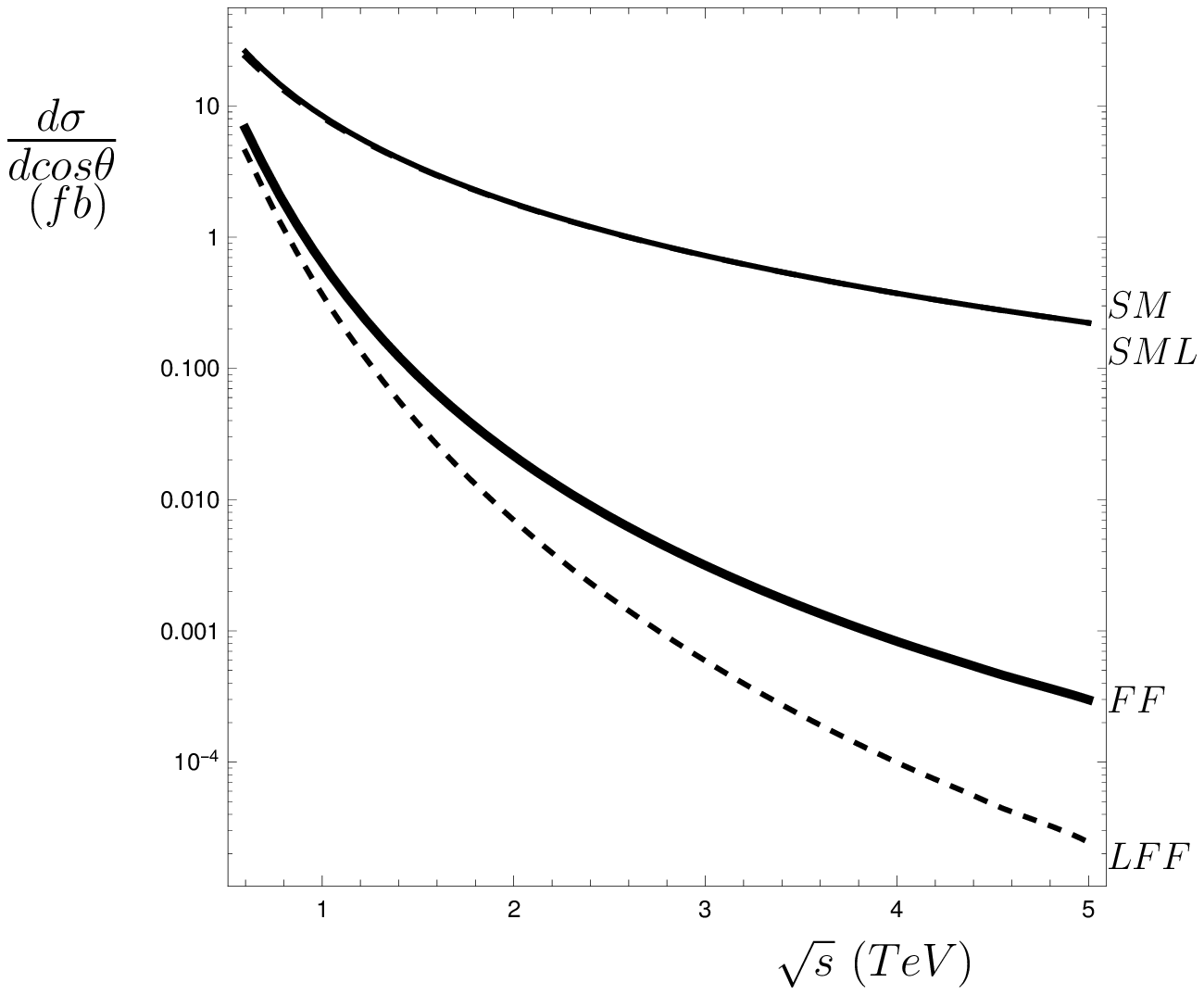, height=8.cm}
\]\\
\[
\epsfig{file=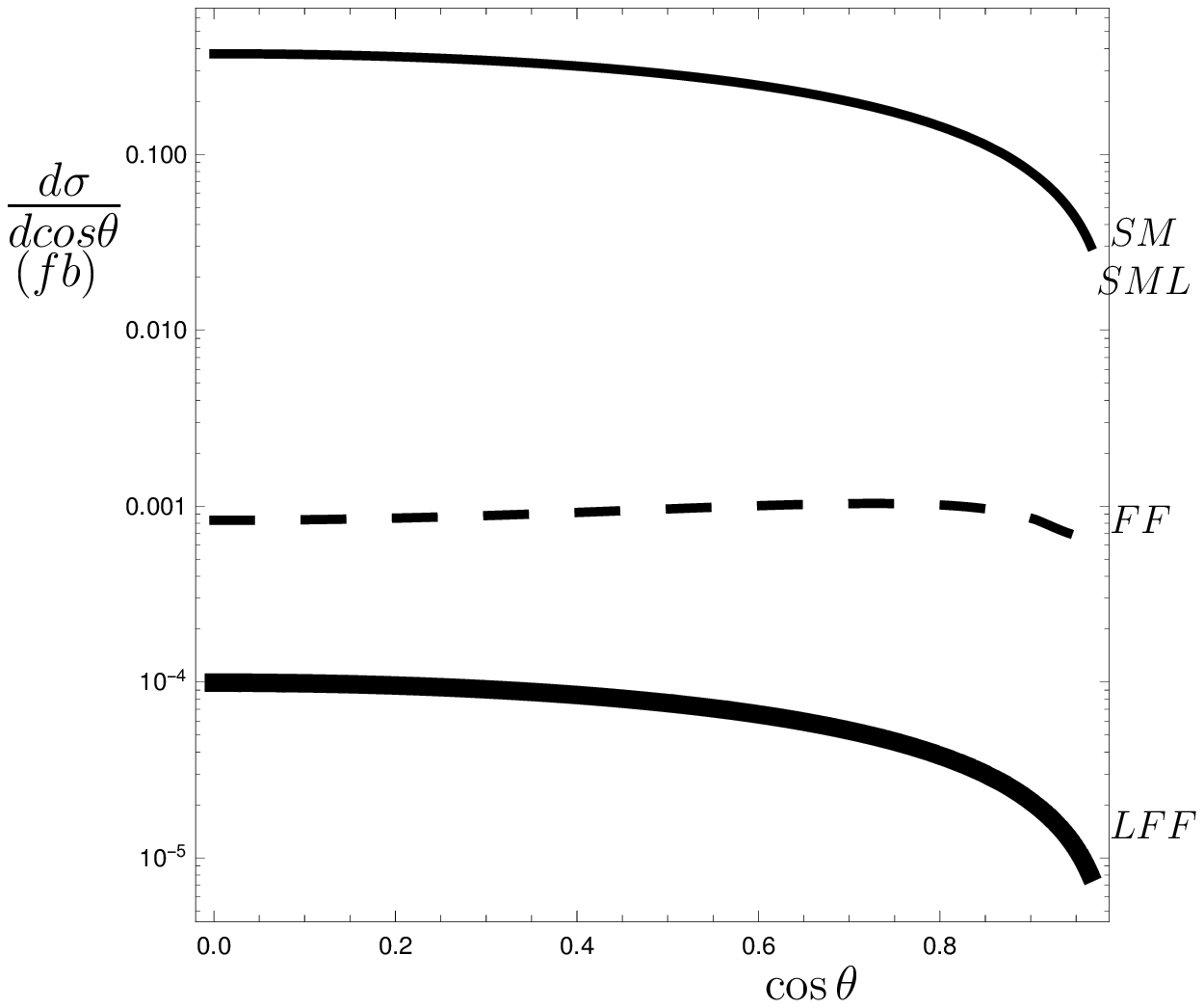, height=8.cm}
\]\\
\vspace{-1cm}
\caption[1] {Energy dependence (upper panel for $\theta=\pi/2$) and angular distribution
(lower panel for $\sqrt{s}=$ 4 TeV)
of the $e^+e^-\to ZH$ cross section. SM refers to the standard
unpolarized case, SML to the standard longitudinal $Z$ production, FF
and LFF to the corresponding cases including the form factor effect.}
\end{figure}

\clearpage

\begin{figure}[p]
\[
\epsfig{file=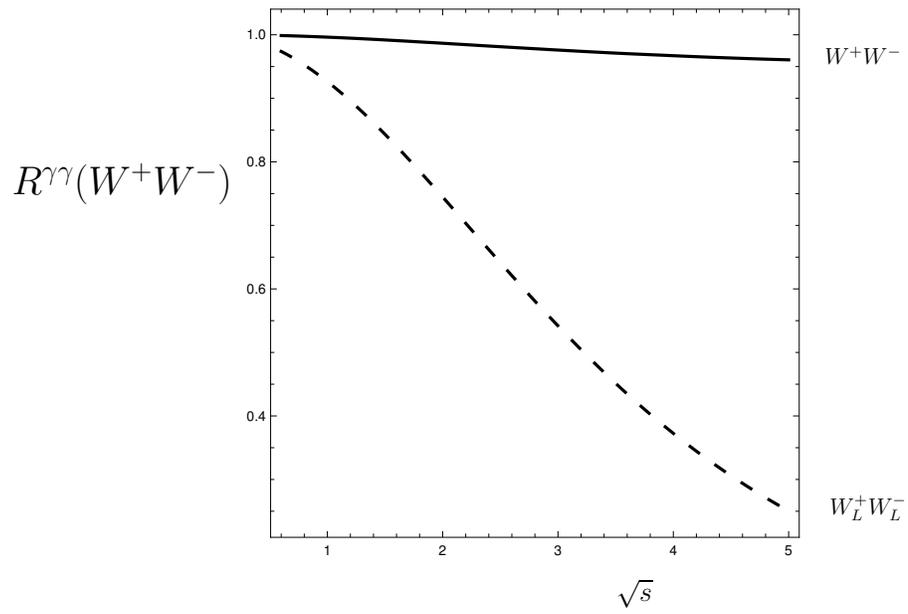, height=8.cm}
\]\\
\vspace{-1cm}
\caption[1] {Ratio of $\gamma\gamma\to W^+W^-$ cross section with form factor
over the standard one in unpolarized case and in $W^+_LW^-_L$ case.}
\end{figure}

\clearpage

\begin{figure}[p]\[
\epsfig{file=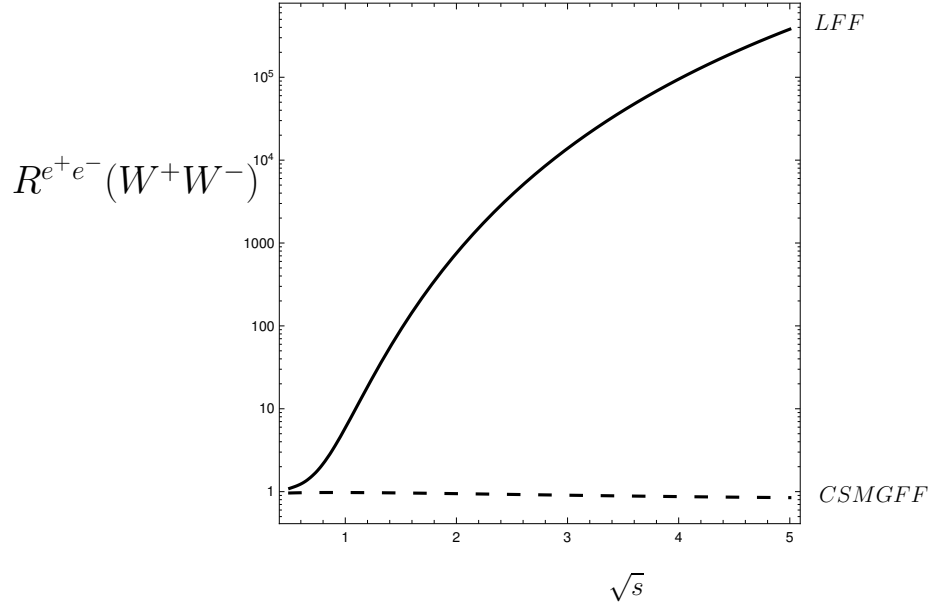, height=8.cm}
\]
\vspace{1.cm}
\[
\epsfig{file=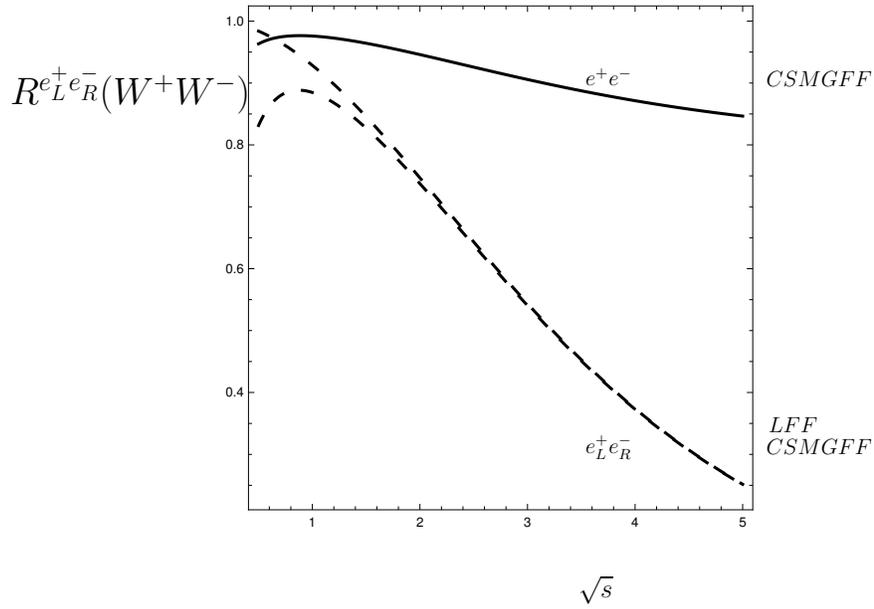, height=8.cm}
\]

\vspace{1.5cm}

\caption[1] {Ratios for $e^+e^- \to W^+W^- $ production;
upper panel for unpolarized $e^+e^-$ with either a crude $W_L$ form factor (LFF)
or with CSM Goldstone equivalence (CSMGFF); lower panel, comparison of the above 
CSMGFF with the case of $e^+_Le^-_R$
polarization and again either a crude $W_L$ form factor (LFF) or with
CSM Goldstone equivalence(CSMGFF).}

\end{figure}

\clearpage

\begin{figure}[p]
\[
\epsfig{file=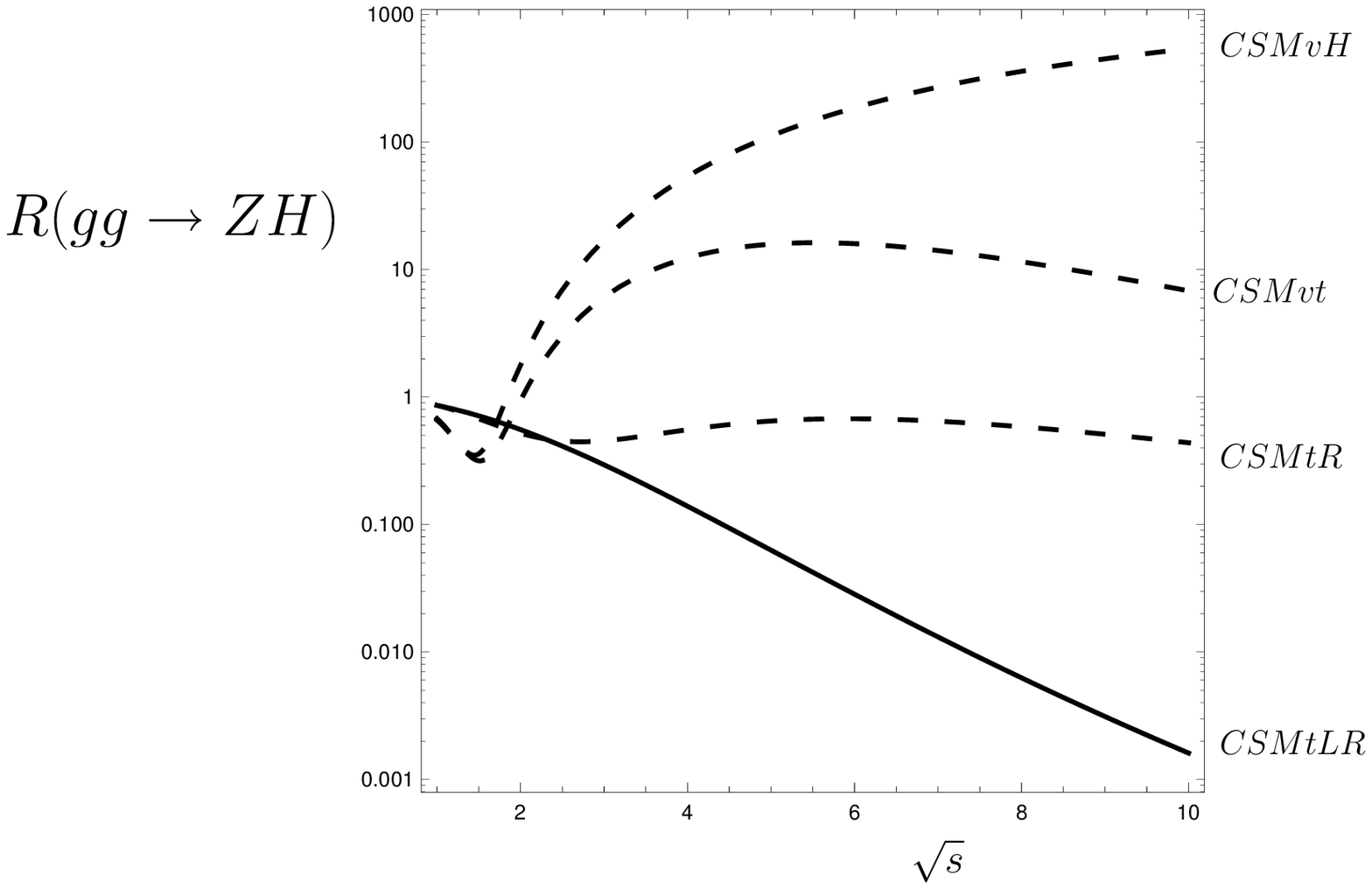, height=8.cm}
\]\\
\vspace{-1cm}
\caption[1] {Ratio of $gg\to ZH$ cross section with form factor
over the standard one.}
\end{figure}

\clearpage

\begin{figure}[p]
\[
\epsfig{file=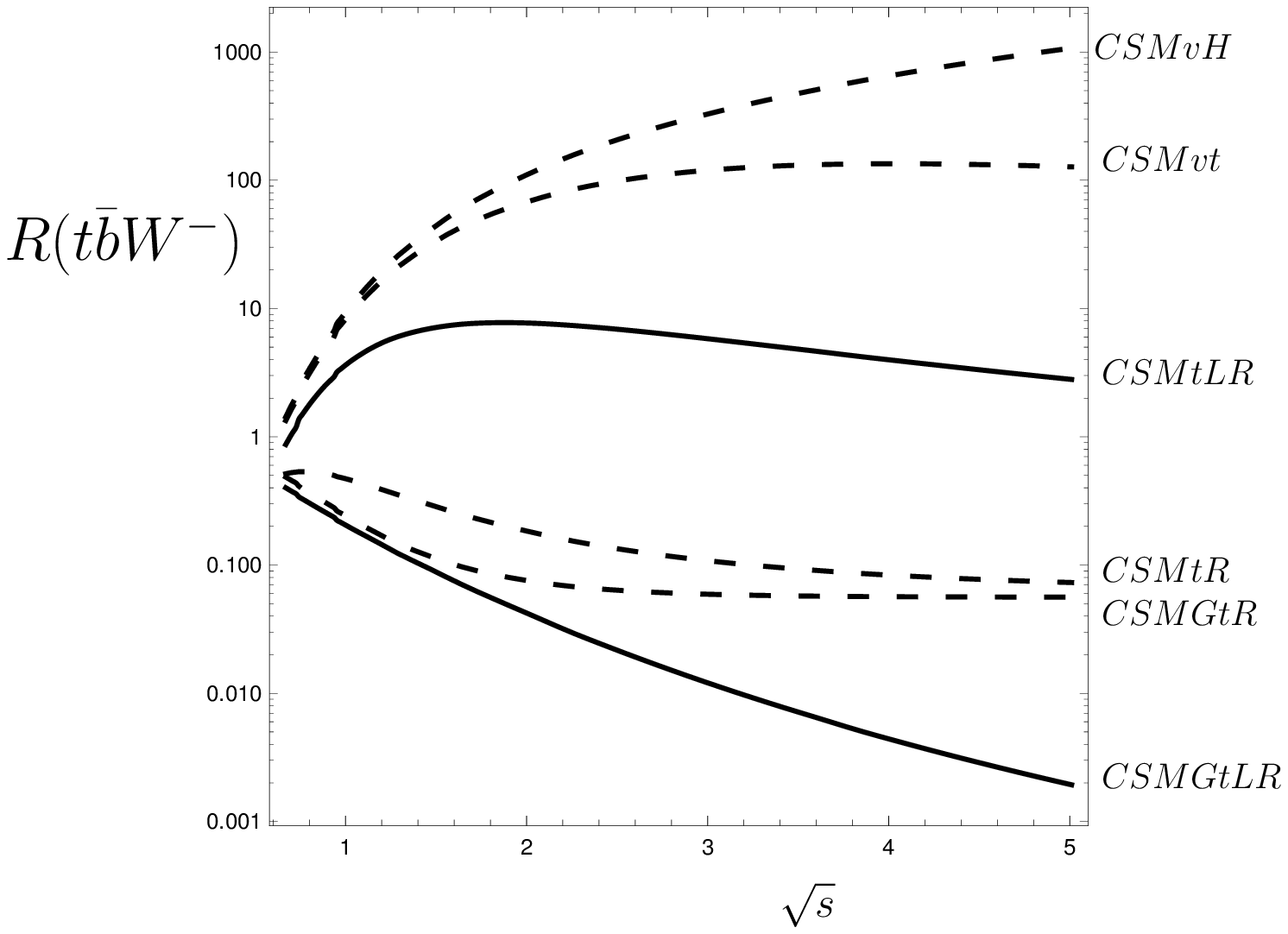, height=8.cm}
\]\\
\vspace{-1cm}
\caption[1] {Ratio of $e^+e^-\to tbW^-$ cross section with form factor
over the standard one.}
\end{figure}

\clearpage

\begin{figure}[p]
\[
\epsfig{file=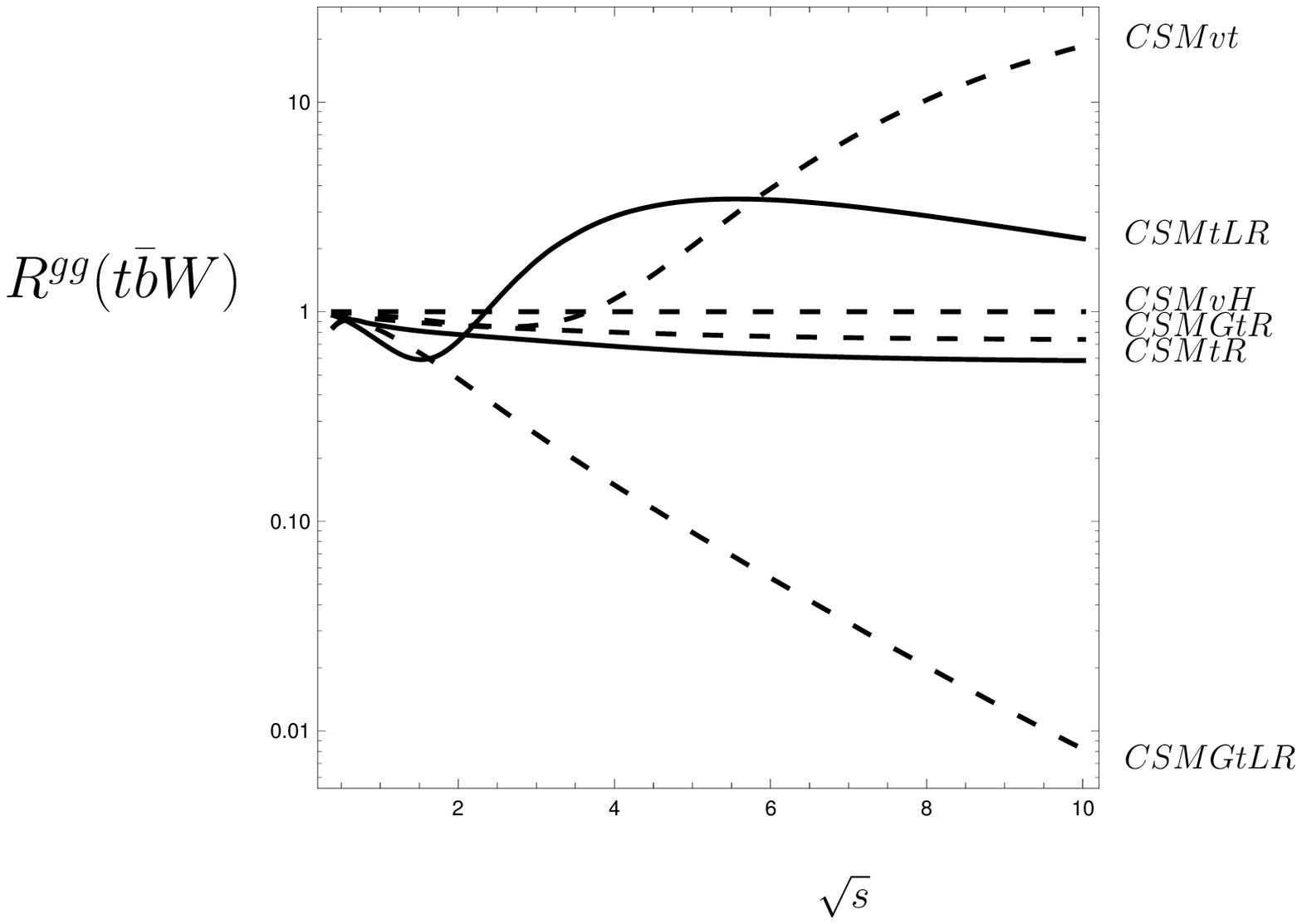, height=8.cm}
\]\\
\vspace{-1cm}
\caption[1] {Ratio of $gg\to tbW$ cross section with form factor over the standard one.}
\end{figure}

\clearpage

\begin{figure}[p]
\[
\epsfig{file=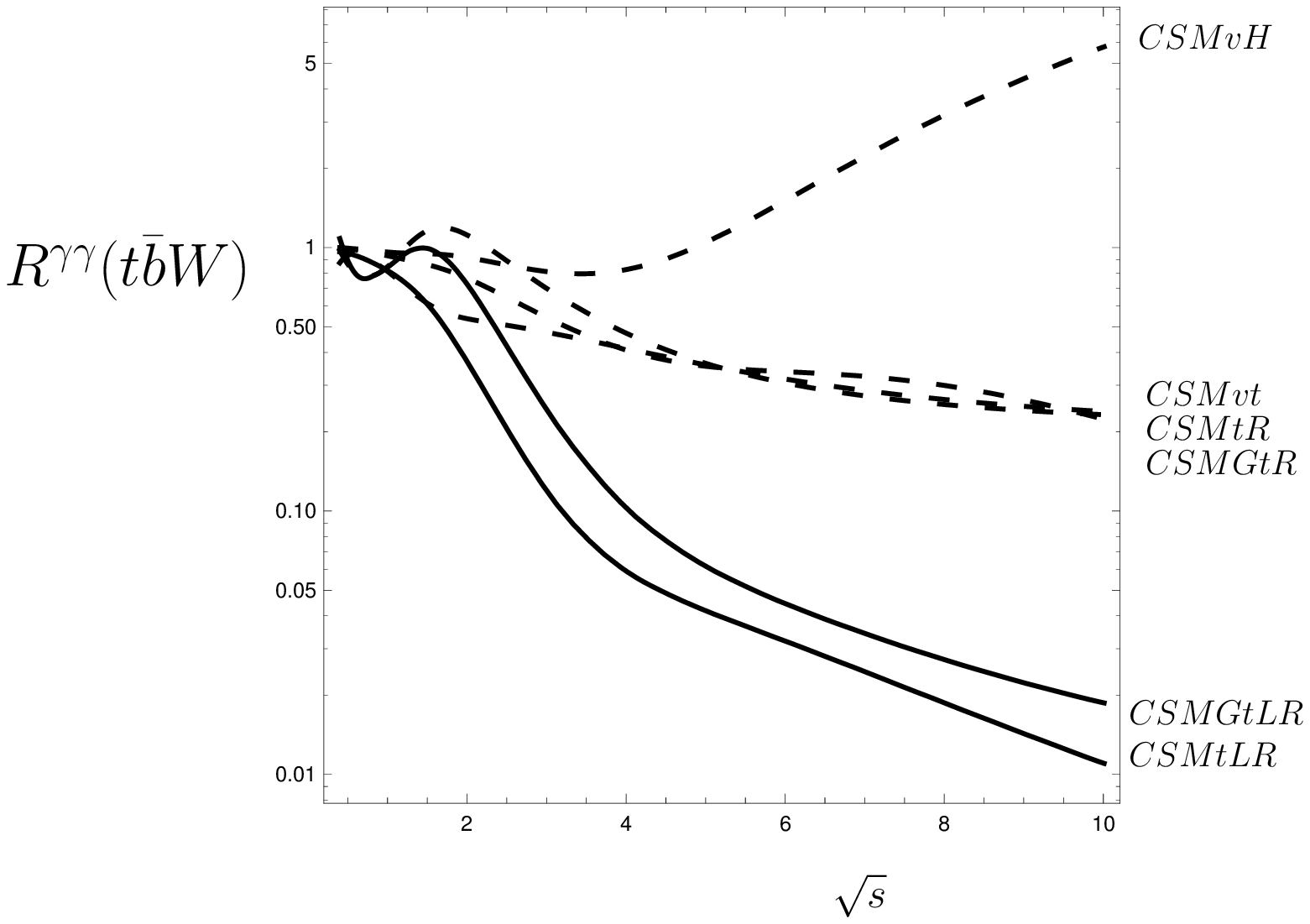, height=8.cm}
\]\\
\vspace{-1cm}
\caption[1] {Ratio of $\gamma\gamma\to tbW^-$ cross section with form factor
over the standard one.}
\end{figure}

\clearpage

\begin{figure}[p]\[
\epsfig{file=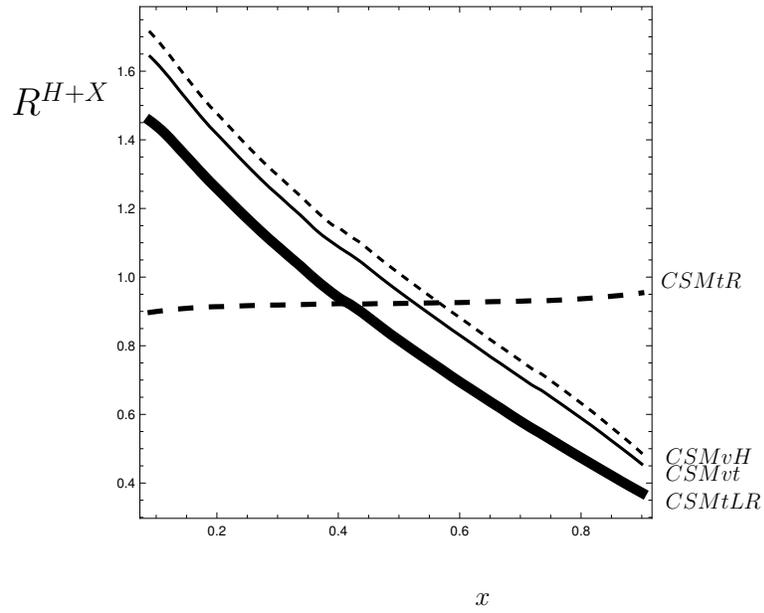, height=8.cm}
\]

\vspace{1.5cm}
\caption[1] {Ratio of $e^+e^- \to H +anything$ with form factor over the standard one.}

\end{figure}

\clearpage

\begin{figure}[p]\[
\epsfig{file=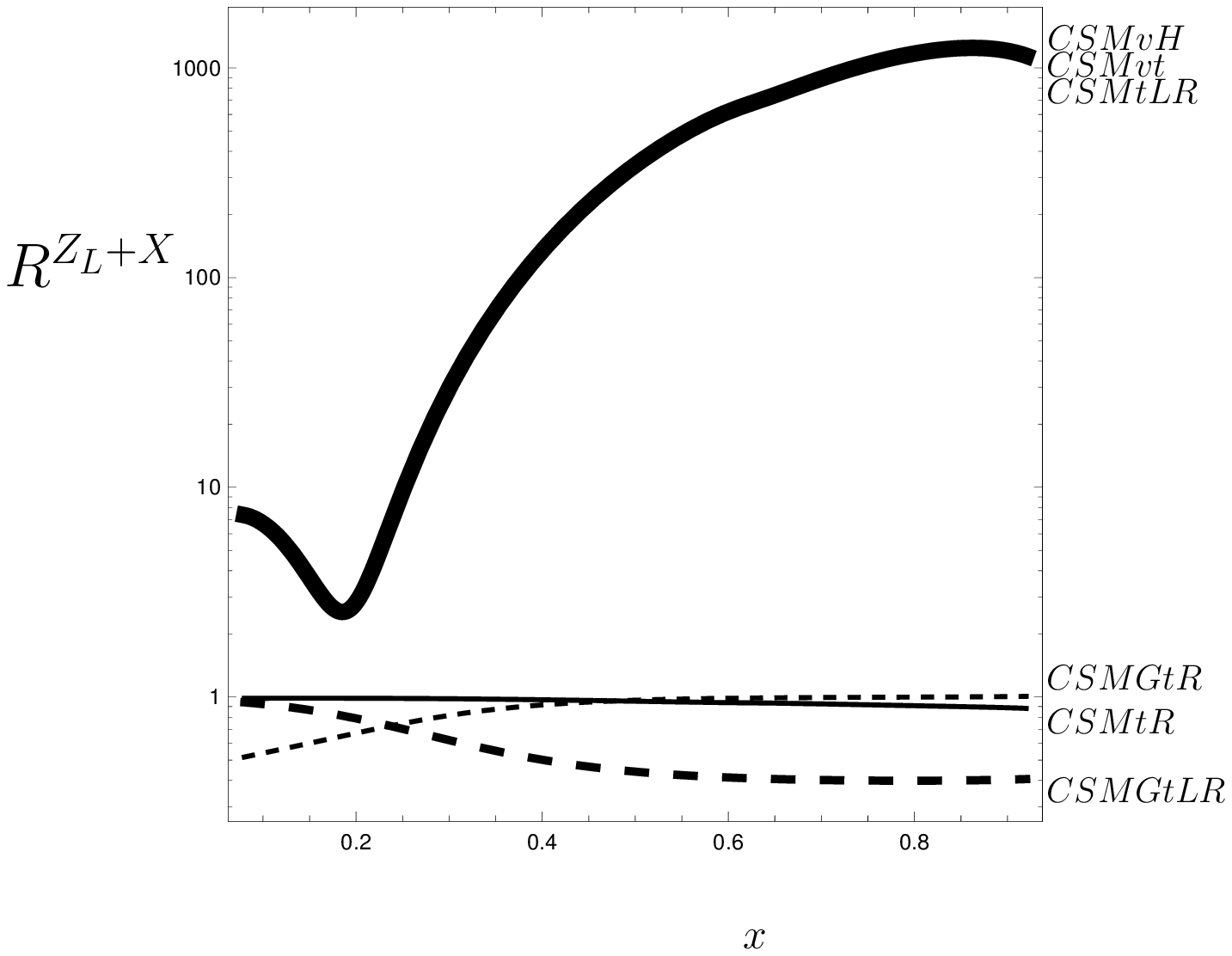, height=8.cm}
\]
\vspace{1.cm}
\[
\epsfig{file=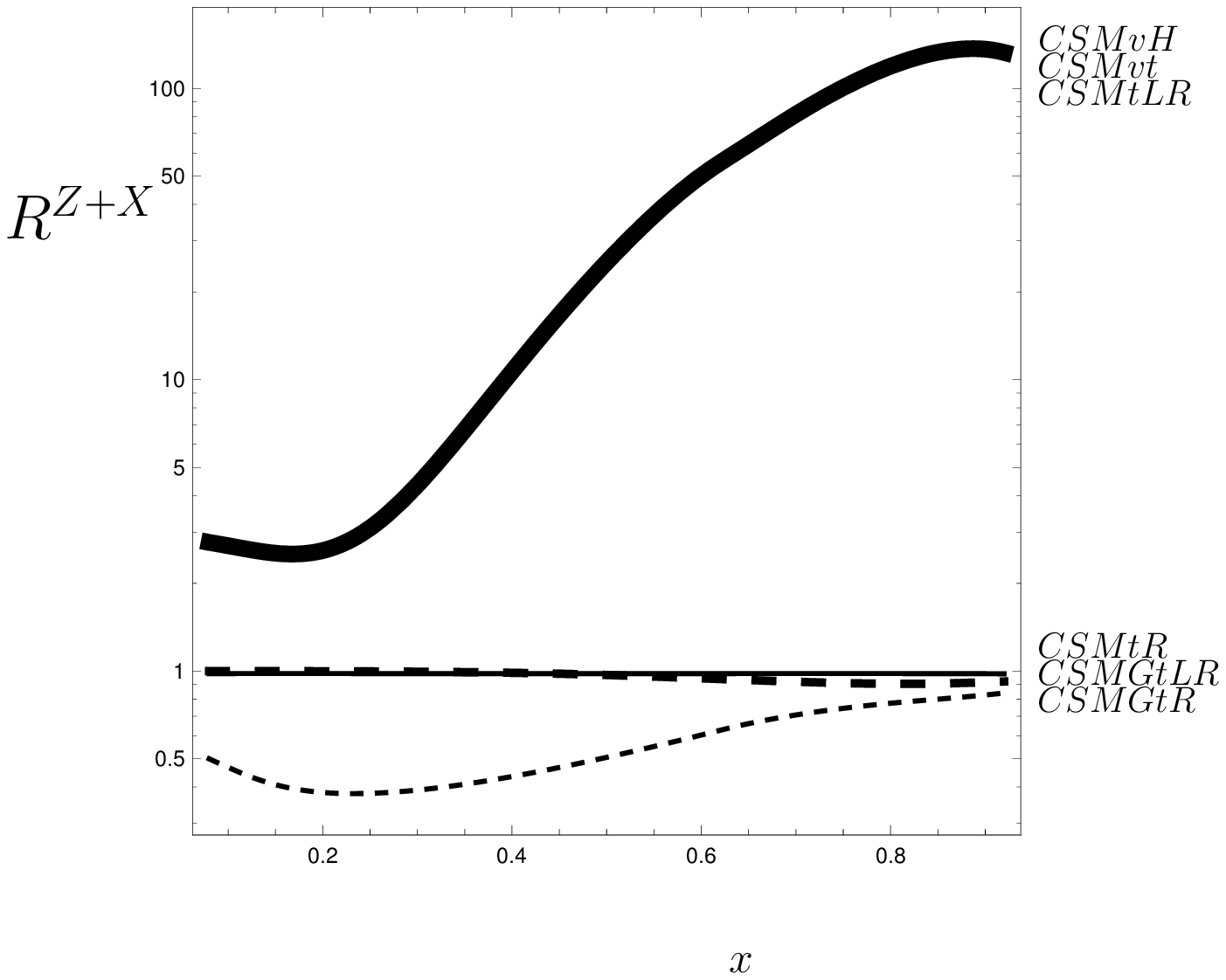, height=8.cm}
\]
\vspace{1.5cm}
\caption[1] {Ratio of $e^+e^- \to Z +anything$ with form factor over the standard one
in the longitudinal and in the unpolarized cases.}

\end{figure}

\clearpage

\begin{figure}[p]\[
\epsfig{file=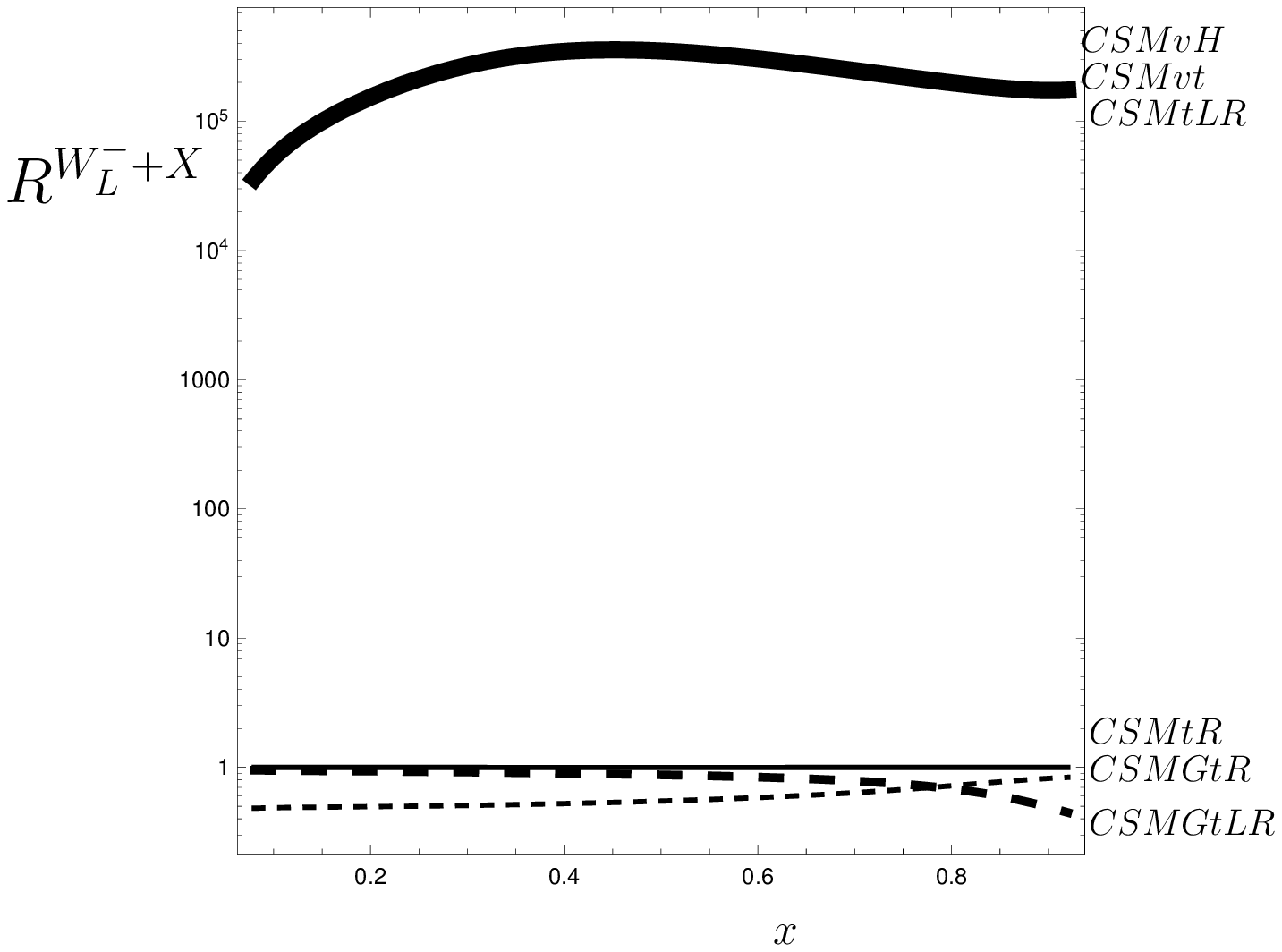, height=8.cm}
\]
\vspace{1.cm}
\[
\epsfig{file=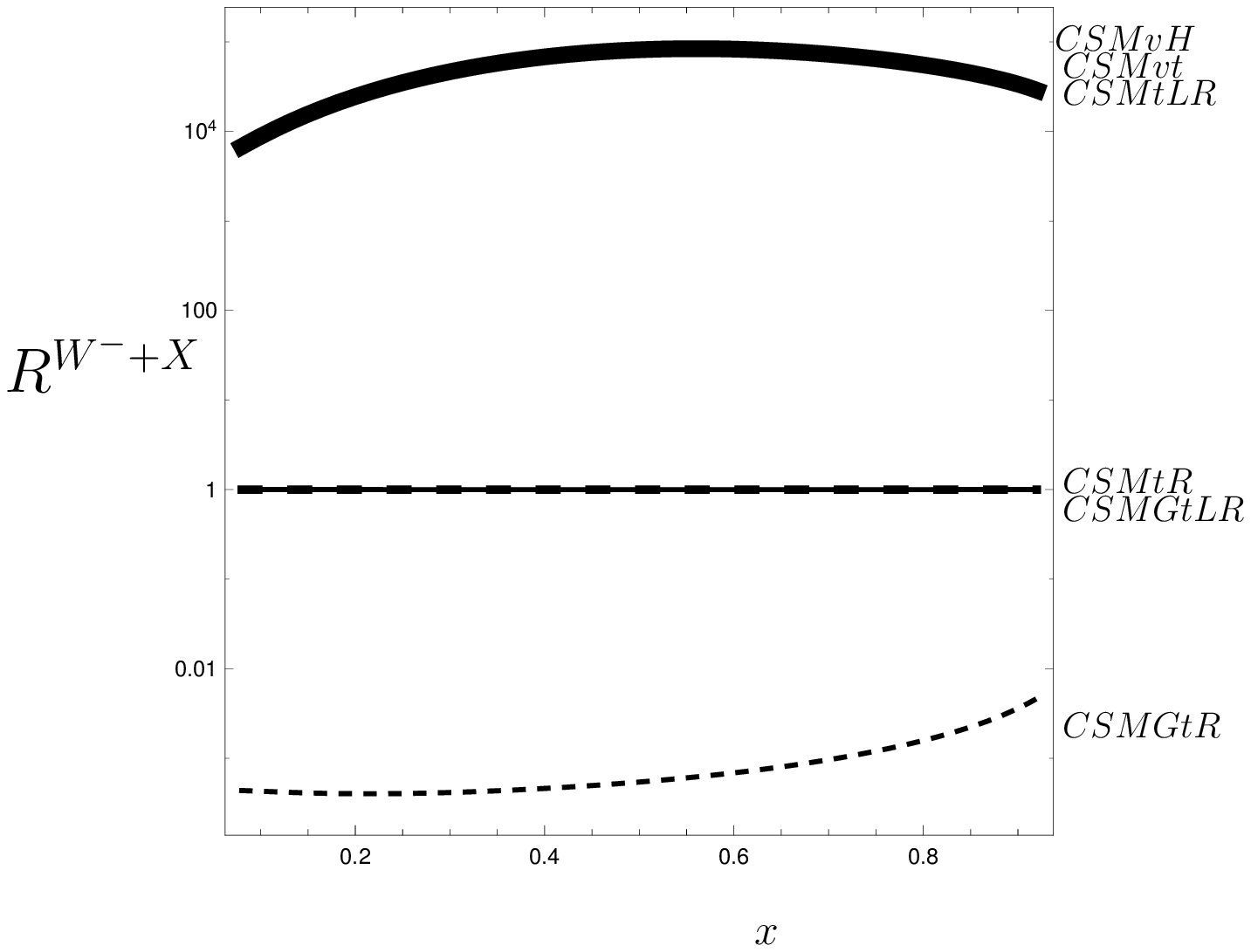, height=8.cm}
\]
\vspace{1.5cm}
\caption[1] {Ratio of $e^+e^- \to W^- +anything$ with form factor over the standard one
in the longitudinal and in the unpolarized cases.}

\end{figure}

\clearpage

\begin{figure}[p]\[
\epsfig{file=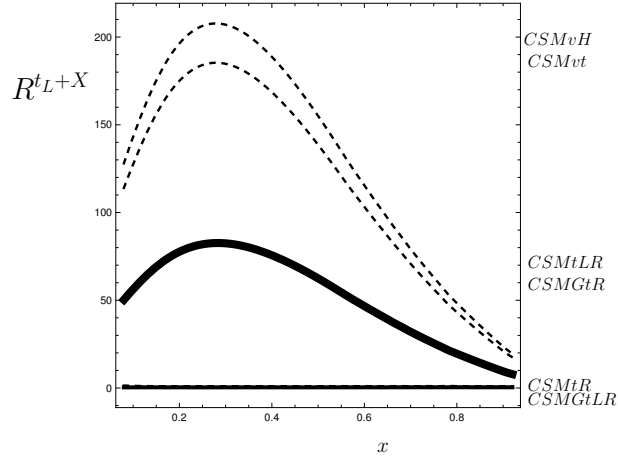, height=6.cm}
\]
\vspace{0.3cm}
\[
\epsfig{file=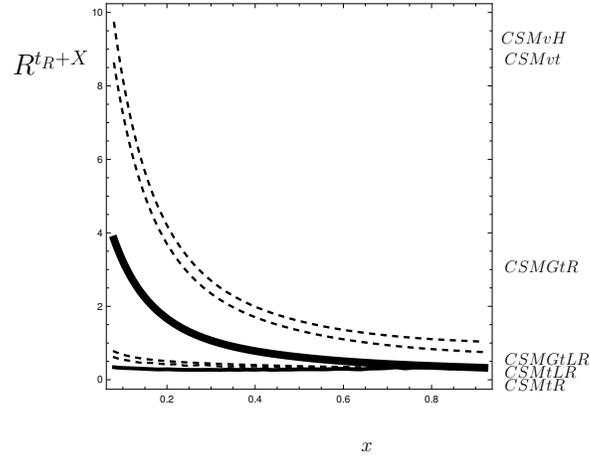, height=6.cm}
\]
\vspace{0.3cm}
\[
\epsfig{file=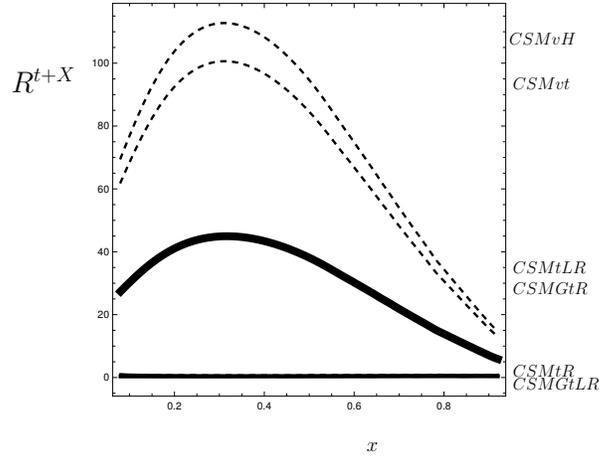, height=6.cm}
\]
\caption[1] {Ratio of $e^+e^- \to t_L +anything$ , $e^+e^- \to t_R +anything$
and $e^+e^- \to t +anything$ with form factors
over the corresponding standard ones.}

\end{figure}

\clearpage

\end{document}